A Single-Molecule Hershey-Chase Experiment

David Van Valen[1*], David Wu[1*], Yi-Ju Chen[2], Hannah Tuson[3], Paul Wiggins[4], and Rob Phillips[1^]

1 Engineering and Applied Sciences, California Institute of Technology, 1200 E. California Blvd,

Pasadena, CA 91125, USA

2 Physics, Mathematics and Astronomy, California Institute of Technology, 1200 E. California Blvd,

Pasadena, CA 91125, USA

3 Department of Biochemistry, University of Wisconsin, Madison, 433 Babcock Drive

Madison, WI 53706 USA

4 Department of Physics, University of Washington, Box 351560

Seattle, WA 98195, USA

* Both authors contributed equally to this work.

^ corresponding author  E-mail: phillips@pboc.caltech.edu, (626) 395-3374

**Summary**

Ever since Hershey and Chase used phages to establish DNA as the carrier of genetic information in 1952, the precise mechanisms of phage DNA translocation have been a mystery [1]. While bulk measurements have set a time scale for *in vivo* DNA translocation during bacteriophage infection, measurements of DNA ejection by single bacteriophages have only been made *in vitro*. Here, we present direct visualization of single bacteriophages infecting individual *Escherichia coli* cells. For bacteriophage lambda, we establish a mean ejection time of roughly 5 minutes with significant cell-to-cell variability, including pausing events. In contrast, corresponding *in vitro* single-molecule ejections take only 10 seconds to reach completion and do not exhibit significant variability. Our data reveal that the velocity of ejection for two different genome lengths collapses onto a single curve. This suggests that *in vivo* ejections are controlled by the amount of DNA ejected, in contrast with *in vitro* DNA ejections, which are governed by the amount of DNA left inside the capsid. This analysis provides evidence against a purely intrastrand repulsion based mechanism, and suggests that cell-internal processes dominate. This provides a picture of the early stages of phage infection and sheds light on the problem of polymer translocation.

**Highlights**

- The infection of *E. coli* cells by single phage λ particles is imaged in real time.
- DNA ejection for phage λ is two orders of magnitude slower *in vivo* than *in vitro*.
- Ejection velocity for different phage mutants scales with the amount of ejected DNA.

**Results**

A schematic of our experimental design is shown in Figure 1A. To visualize DNA translocation, we use a fluorescent marker to stain the viral DNA while it still remains in the capsid [2]. Phages are first incubated in the appropriate DNA stain before dialyzing away excess dye. The stained phages are then briefly bound to bacterial cells, which are pipetted into a flow chamber. The sample is then washed with buffer and then imaged with brightfield and fluorescence microscopy over a sufficiently long time to monitor the infection process. The signature of an ejection event is a loss of fluorescence in the virus and a concomitant increase in the fluorescence within the bacterial cell (Fig. 1B). We were encouraged that such an experiment was possible from a number of clues in the literature. First, phages stained with cyanine dyes have previously been observed to infect cells [3]. Second, studies characterizing the interaction of cyanine dyes with DNA inside the phage capsid have demonstrated that for some dyes, stained phages remain intact [2]. Cyanine dyes have also been used to study the kinetics of viruses in living eukaryotic cells, demonstrating that some dyes have limited cytotoxicity [4]. To identify a suitable dye, we screened a number of candidate dyes for their ability to (i) penetrate the phage capsid (Fig. S1), (ii) maintain phage stability and infectivity while stained (Fig S2A-E), (iii) preserve *in vitro* ejection kinetics while stained (Fig. S3A-C), (iv) not cross the membrane of living cells (Fig. S2F), and (v) be sufficiently bright for quantitative analysis (Fig. S4). Our screen identified SYTOX Orange as a dye with all of these properties; these controls are described in full detail in the Supplemental Information.

**Single-Cell DNA Ejection Trajectories**

A typical *in vivo* ejection event for phage λcI60, a strain with a wild-type genome length of 48.5 kbp, obtained from this assay is shown in Figure 2. As seen in Figure 2A, the attachment of the viruses to the host is revealed by the presence of diffraction-limited spots on the cell surface. We identified pixels associated with either the virus or the cell (Fig. 2B) and queried the fluorescence intensity as a function of time. As shown in the montage of images (Fig. 2C), the ejection process is characterized by a loss of fluorescence intensity in the phage and a concomitant increase in fluorescence in the cellular interior (Fig. 2D, representative Movies S1, S2). We note that the fluorescence inside the cell is diffuse; this reflects the dye molecules unbinding from the phage DNA and redistributing themselves along the host genome. This is expected from the residence time of SYTOX Orange on DNA, which is ~ 1s [5]; we verified this with buffer exchange experiments (Fig. S3B). In the particular trajectory shown, the increase in cellular fluorescence is roughly equal to the decrease in phage fluorescence. The decrease in signal at the end of the trajectories in Figure 2D can be accounted for by photobleaching (Fig. S3C).

**Single phage *in vivo* ejections are two orders of magnitude slower than *in vitro* and display pausing**

The results of a number of distinct ejection events for λcI60 are shown in Figure 3. For the measurements shown here, the fluorescence signal associated with the virus decreases on a time scale of minutes, a factor of 10 to 100 times longer than the corresponding dynamics observed *in vitro* [7,10]. We define the ejection time as the time required for 80% of the fluorescence intensity to leave the viral capsid. The mean time for ejection for λcI60 and the corresponding standard deviation was 5.2 ± 4.2 minutes (n = 45). In addition to the widespread variability in ejection time, there are a number of ejections which demonstrate pausing events. We define a pause as a non-decreasing fluorescence level greater than 2 minutes (our time resolution was typically 1 minute, so 2 minutes is the minimum time required to rule out spurious events); the mean pause time for λcI60 was 5.4 ± 4.1 minutes (n = 14). Based on this observation, we have partitioned the trajectories into two classes: single-step and paused. These are shown in Figure 3A and Figure 3B, respectively. A sample paused ejection is shown in Movie S3. The full set of trajectories for λcI60 are on-line at the corresponding author's website.

We next asked if a reduction in driving force would produce significant differences in DNA translocation rates, an idea already used in our earlier *in vitro* measurements [8,10]. In this earlier work, the ejection of phage λcI60, which has a genome length of 48.5 kbp was compared to the ejection of phage strain λb221, which has the b region of the genome removed and a shorter genome length of 37.7 kbp. Through bulk and single-molecule *in vitro* experiments, it was shown that the amount of DNA inside the viral capsid was a control parameter for *in vitro* DNA ejection [8,10]. Once phage λcI60 has ejected 10.8 kbp of DNA, the ejection forces and dynamics are equivalent to that of phage λb221. To explore the effect of genome length changes on DNA translocation rates *in vivo*, we performed our *in vivo* ejection assay for phage λb221. The mean time for ejection was 2.58 ± 2.34 minutes (n=18). One paused ejection was also observed, with a pause time of 5 minutes. For λb221, we also observed a number of ejections (n=10) that did not finish during the course of the movie. Stalled ejections were not observed for λcI60. One possibility is that stalling events are related to λb221's shorter genome and the consequent loss of a potential binding site that assists in entry. Stalled ejections were not included in the averages given earlier. The full set of trajectories for phage λb221 are on-line at the corresponding author's website.

**An ensemble view of *in vivo* ejection shows the amount of DNA in the viral capsid is not the governing control parameter**

Our measurements on both the wild-type and shortened genomes provide an opportunity to quantitatively examine the DNA translocation kinetics. One quantity of interest is the first-passage time for ejection, defined as the time at which a certain amount of DNA has entered the cell. For each trajectory, we extracted the first passage time for 20%, 50%, and 80% completion, as determined by the decrease in the starting phage fluorescence. The first-passage time distributions for λcI60 and λb221 are shown in Figure 4A. By taking the mean of this distribution, we obtain the mean first-passage time; this is plotted in Figure 4B. We note that the mean first-passage time is a quantity that is amenable to theoretical calculations [9]. Another way to view the mean first-passage time is as an "average" ejection trajectory. When viewed in this way, one interpretation of Figure 4B is that, within the error of the measurement, the "average" trajectories for λcI60 and λb221 have considerable overlap. For both λcI60 and λb221, the velocities plateau after ~50% of the genome length as shown in Figure 4C and 4D.

Another method of analyzing the dynamics is to determine the mean velocity at different amounts of DNA remaining in the capsid during an ejection. In this analysis, the velocity at each landmark is extracted from each trajectory. The mean velocity is then recorded and plotted as a function of the amount of DNA remaining in the capsid. The result of such an analysis for λcI60 and λb221 is shown in Figure 4C. This plot shows that there is little overlap between the two curves and that for lower amounts of DNA remaining in the capsid, the velocity for λb221 is higher than λcI60. This is to be contrasted with *in vitro* measurements where there is significant overlap between the two curves, with the "data collapse" in that case signifying that the dynamics are equivalent after λcI60 has ejected its first 10.8 kbp [10]. An alternative way to visualize this same data is to plot the mean velocity versus the amount of DNA ejected into the cell; this is shown in Figure 4D. When plotted in this fashion, there is considerable overlap between the two curves, with a small difference observed for the first 20 kbp of ejection. This analysis is consistent with the mean first passage time analysis, which showed considerable overlap when the first-passage time was also plotted against the amount of DNA ejected.

**Discussion**

For bacterial viruses, genome delivery is at the heart of the viral life cycle. And yet, this critical process of transfer of viral DNA from the virus to its host remains enigmatic as does the *in vivo* process of polymer translocation more generally. Beyond a purely intellectual understanding of this process, phage-mediated horizontal transfer of virulence factors is a noted cause of world-wide dysentery [11,12], antibiotic resistance [13], and is a key player in the evolution of geologic bacteria communities [14]. Our objective was to design and perform an experiment with single-molecule resolution that would permit us to watch this process in real time, to measure the specific functional form of the speed of ejection as a function of the amount of DNA left in the capsid (and hence as a function of the amount out as well) and to characterize the cell-to-cell variability of that process; we accomplished this for both wild type lambda phage (48.5 kbp) and a mutant with a shorter genome (37.7 kbp) using a fluorescent staining strategy in conjunction with time-lapse microscopy. These experiments reveal that the DNA translocation process is subject to strong cell-to-cell variability with the ejection times exhibiting a wide range from approximately 1 to 20 minutes. A number of ejections exhibited pauses with some never reaching completion during the course of the experiment. Our single-molecule measurements are consistent with earlier estimates of a minute time scale for *in vivo* genome delivery of phage lambda from bulk experiments [15,16], as the bulk data should only be compared to the fastest observed single-molecule ejections, since Southern blot analysis would pick out the earliest infections, but not the entire distribution of infection times.

A number of different hypotheses have been formulated for the actual translocation mechanism for phage λ. In addition to the driving force due to the packaged DNA, these models propose that thermal fluctuations, hydrodynamic drag, and active molecular motors might each play a role in bringing the viral DNA into the bacterial cell [6,9,16,18,19]. With respect to identifying the correct ejection mechanism for phage λ, our results provide both surprises and useful insights that constrain the space of possible models and will guide future modeling efforts. One key result is that the length of DNA remaining inside the capsid is not the sole control parameter that governs the ejection dynamics, as it is *in vitro*. In the *in vitro* experiments, the approximate collapse of the data from the different genome lengths on a single curve revealed that the DNA within the capsid is driving the kinetics of ejection [8,10]. By way of contrast, in the *in vivo* ejection experiments reported here, an approximate data collapse is only revealed when the velocity is plotted with respect to how much DNA is out of the capsid and in the cell rather than how much DNA remains within the capsid. No collapse is seen when the velocity is plotted against the amount of DNA remaining inside the capsid.

The lack of data collapse with respect to DNA left in the capsid has significant implications for the role the energy stored in the compacted DNA plays during the *in vivo* ejection. If some significant portion of the ejection process were governed solely by the energy in the compacted DNA, then during that portion we would expect the dynamics of λcI60 and λb221 to be identical when the amount of DNA remaining in the capsid is identical. This is the *in vitro* case as studied in [8,10]. Because the DNA-DNA repulsion inside the capsid is highest when the capsid contains more DNA, such a period would likely be at the beginning of the ejection process. As seen in Figure 4C, however, there exists no period of overlap between the velocity curves for the two phage strains, and hence no period during the ejection process where the length of DNA in the capsid is the sole control parameter. From this observation, we can conclude that there is no period of time during which *in vivo* DNA ejection is governed solely by the intrastrand repulsion inside the capsid. Two-step models in which the first half of the genome is delivered by the energy stored in the compacted DNA and the remainder is delivered by another mechanism are also not consistent with our data.

Another consequence of the data collapse is the possibility that the amount of DNA ejected, as opposed to the amount of DNA in the capsid, is a key control parameter for this system. This picture would be consistent with models in which the mechanism is internal to the cell, as the only information such a mechanism would utilize is the amount of DNA that has been brought inside the cell. The collapse of data on to a single curve is a powerful argument that has been used before to identify control parameters for *in vitro* DNA ejection as well as the lysis-lysogeny decision [8,10,20]. One limitation to applying this argument is that only two genome lengths have been tested here, and it is possible that this collapse does not hold for all viable genome lengths. Such reasoning also does not exclude a mixed picture, as mentioned above.

The origin of the apparent pauses might provide information about the ejection mechanism, as DNA-based motors acting against a load have been observed to pause [21,22]. However, the pauses observed here are much longer than the pauses observed for motors and it is possible they could simply be a reflection of the cell-to-cell variability in turgor pressure, as when intracapsid pressure equalizes with turgor pressure, there is no net driving force. Post-pause resuming of phage DNA entry could thus indicate a secondary mechanism in conjunction with pressure, for high turgor cells. Another possibility is that the pauses observed here might also be related to mechanisms proposed for pauses observed *in vitro* for phage T5 [23,24]. However, this is unlikely as pauses are not observed for phage lambda *in vitro* [7,10]. We also cannot rule out the possibility that SYTOX Orange intermittently interferes with the ejection process. Another possibility is that the DNA in the phages are nicked; however, we never see any pausing in *in vitro* experiments done from the same phage lysate.

We note that our results are contrary to what was shown in T7, in which a constant DNA ejection rate was seen with bulk measurements [16]. T7 has a capsid similar in size to lambda (60 nm vs. 58 nm, respectively) with a 40 kbp genome; however its tail is considerably shorter (23 nm vs. 150 nm, respectively) [25,26]. It has been suggested that a constant velocity is suggestive of a purely enzyme driven model such as a molecular motor [16]. Such a feature is not seen in our data, as Figure 4D shows that once ~ 20 kbp of DNA has been ejected, there is a marked decrease in the ejection velocity. However, the non-linear force-velocity relationship seen *in vitro* and the presence of pN level forces from the DNA-DNA repulsion inside the capsid make it unclear whether a constant ejection rate prediction would be true for lambda.

The current data does not match previous calculations on mechanisms based on DNA binding proteins and thermal fluctuations [8]. Those calculations predict that after the first 10.8 kbp of DNA from λcI60 has been ejected, it should have the same dynamics as λb221 which, as discussed earlier, is not consistent with our data. Also perplexing is the time scale of ejection. The origin of the friction that sets the timescale for ejection is poorly understood, both *in vitro* and *in vivo.* A number of models assume a linear relationship between force and velocity, but it is now known that this assumption is not true *in vitro* [9,10,27]; we suspect it is not true *in vivo* either.  A better understanding of this friction would allow the current theories to speak quantitatively about the experimental observables in the single-molecule *in vitro* and *in vivo* DNA ejection experiments.

In summary, we have examined the DNA ejection process for bacteriophage λ *in vivo* at the single-molecule level. We note that the techniques explored in this work may be generalizable to the study of other bacteriophages. It would be especially interesting to see a comparison between the bulk and single-molecule dynamics for bacteriophage T7 as bulk experiments have shown that the speed is constant throughout the ejection process *in vivo* [16,25], as opposed to the variable rate reported here. We also note that the experimental platform presented here can be used to explore the effects of various genetic, chemical, and mechanical perturbations on the ejection process. Altogether, this work provides quantitative data that constrains the class of models that can be used to explain the mechanism behind genome delivery in bacteriophage λ and provides insight into the more general problem of polymer translocation.

**Experimental Procedures**

**Real-time imaging of DNA ejection** *in vivo*. Glass coverslips were cleaned by sonication for 30 minutes in 1 M KOH followed by sonication in 100% ethanol with copious rinsing with purified water in between, and then dried on a hot plate. The coverslips were then briefly (5 seconds) immersed in a fresh solution of 1% polyethyleneimine, transferred into purified water, and finally dried with a stream of air. A microscope slide, double-sided tape, and the treated coverslip were then assembled into a flow chamber. *E. coli* strain LE392 was grown up overnight in LB media at 37 °C (see SI for all buffer formulations). The saturated culture was then diluted 1:100 in M9 maltose-sup and grown for 3 hours at 37 °C until the culture reached OD600 ~ 0.3. Plate lysate of the desired phage strain (see Supplemental Information) was centrifuged for 5 minutes at 13,000 g to remove bacterial debris. The supernatant was recovered and then stained with SYTOX Orange at a final concentration of 500 nM for 3 hours at room temperature. Prior to binding stained phages to *E. coli*, free dye was removed by diluting 100 µL of the phage suspension and then centrifuging the sample across a 100 kDa (EMD Millipore,UFC910008) filter 4 times. Each round of centrifugation led to a 40-fold dilution of dye, reducing the final free concentration of dye to less than 200 pM. After the final round, the phages were brought up to the original volume of 100 µL with M9sup. Phages were then bound to cells by mixing ~ 50 µL of cells with phage at a multiplicity of infection (MOI) of ~ 7 for 1 minute at room temperature. The cells were then flowed into the flow chamber and allowed to adhere to the surface for 2 minutes at room temperature. Occasionally, phages were bound to cells by mixing 10 µL of cells with phage at a MOI of ~ 1-5 and incubating on ice for 30 minutes. Incubation of the cells in the flow chamber took place on ice as well. The initiation of phage ejection is slowed down considerably within this time period in either of these two conditions [20,28,29].

After the incubation, the flow chamber was washed with 200 µL of M9sup with 1% GODCAT mixture, 1% beta-mercaptoethanol, and 0.5% glucose. The chamber was then sealed with valap and imaged on a Nikon Ti-E Perfect Focus microscope using a mercury lamp and TRITC filter (Semrock, LF561) set at 37 $^{o}$C. Snapshots of both the phase and fluorescence channels were taken either 1 or 4 times a minute, with a fluorescence excitation time of 500 ms or 300 ms, respectively. Images were collected using either a Hamamatsu C8484 camera, a Photometrics CoolSNAP ES2 camera, or an Andor iXON EMCCD camera. We observed better conservation of fluorescence between the phage and the cell with the Hamamatsu and Photometrics cameras as opposed to the Andor camera. The EM gain of the Andor camera allowed for shorter exposure times and higher time resolution.

**Acknowledgements**

We are grateful to a number of people for help with experiments, advice, and critical commentary on the manuscript including: Heun Jin Lee, Maja Bialecka, Phillips lab, Talia Weiss, Vilawain Fernandes, Kari Barlan, Paul Grayson, Ido Golding, Lanying Zeng, Bill Gelbart, Chuck Knobler, Francois St. Pierre, and Drew Endy. We are also grateful to Ron Vale, Tim Mitchison, Dyche Mullins, and Clare Waterman as well as several generations of students from the MBL Physiology Course where this work has been developed over several summers. We also gratefully acknowledge financial support from several sources, including a NIH Medical Scientist Training Program Fellowship, a Yaser Abu-Mostafa Hertz Fellowship, and a NIH Director's Pioneer Award. We also acknowledge the support of NSF grant number 0758343.

**Figures**

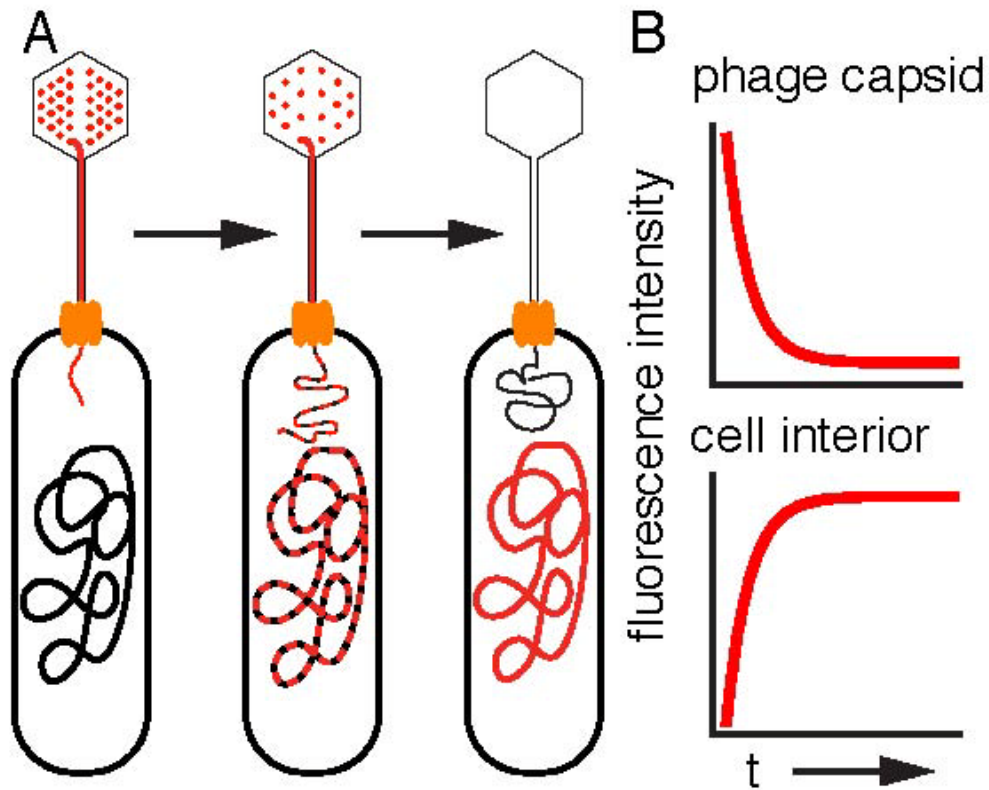

**Figure 1. A schematic for monitoring DNA translocation with pre-ejection labeling.** A. The DNA is stained while still in the capsid. During ejection, the phage DNA carries its complement of cyanine dye with it, transferring fluorescence intensity from the virus to the cellular interior. Eventually, the dye falls off the phage DNA and rebinds to the bacterium's genome. B. The timing of ejection is determined by measuring the loss of fluorescence intensity from the capsid; the concomitant increase in intensity in the cellular interior serves to verify that phage DNA has entered the cell.

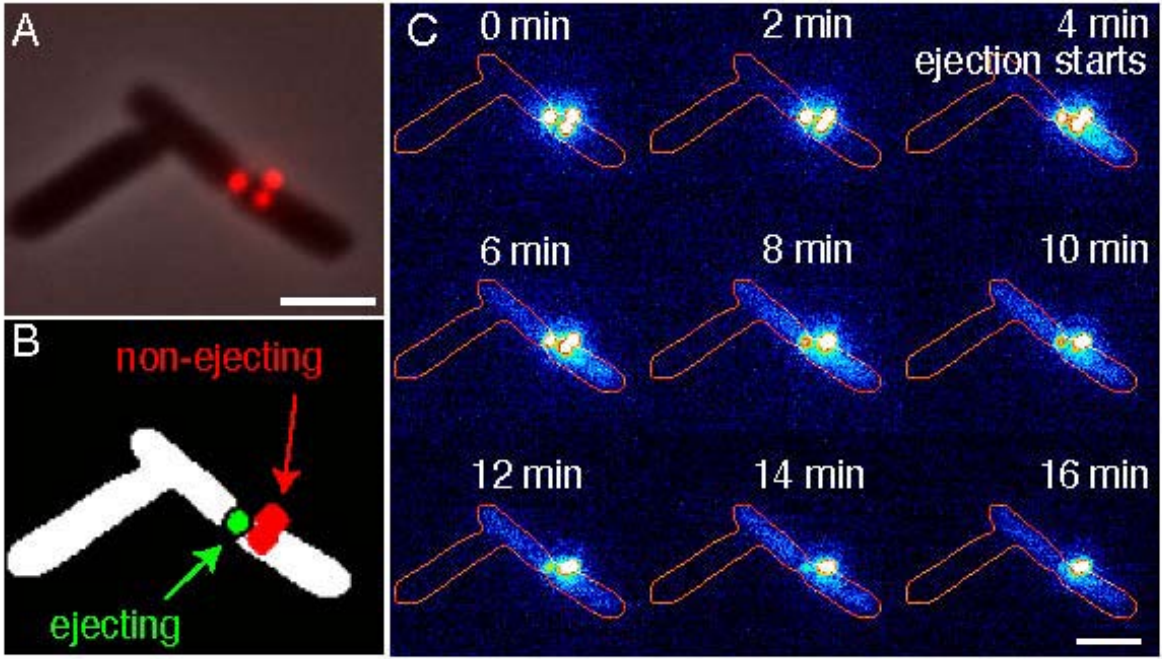

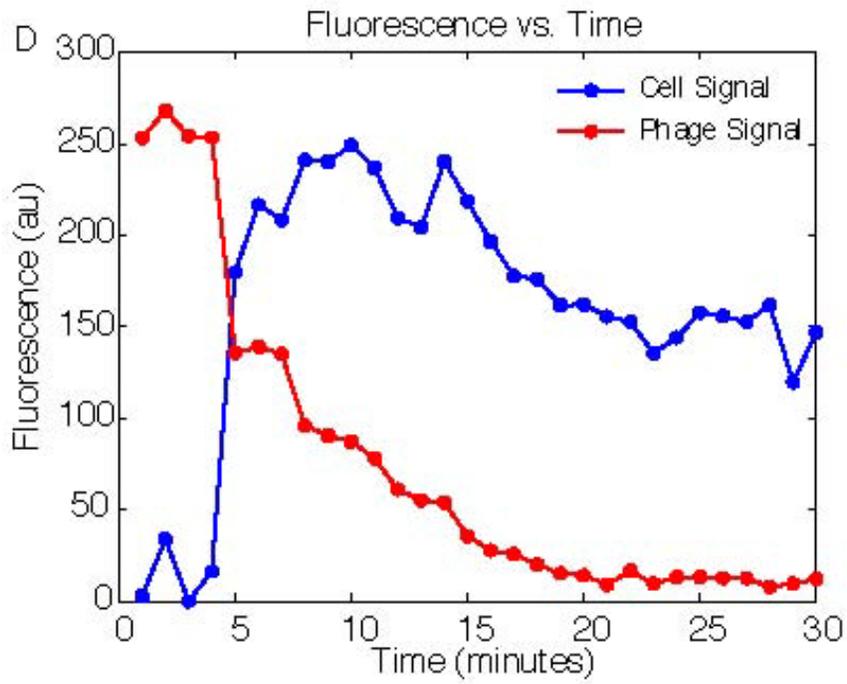

**Figure 2. Dynamics of DNA ejection.** A. Viruses attached to the cell surface in this fluorescence image merged with its brightfield counterpart. B. Segmentation masks of the cell (white), the phage which ejects its DNA (green), and the phages that do not eject their DNA (red). C. Time sequence of the fluorescence in the cell. The edge of the cell is outlined for reference. D. Fluorescence intensity as a function of time. The intensity of the phage segmented region and the cell segmented region are each plotted separately. The fluorescence intensity inside the non-ejecting phage mask is stable; this is shown in Figure S4A. Note that in this ejection there appear to be steps and pauses. The scale bar in (A) and (C) is 2 microns.

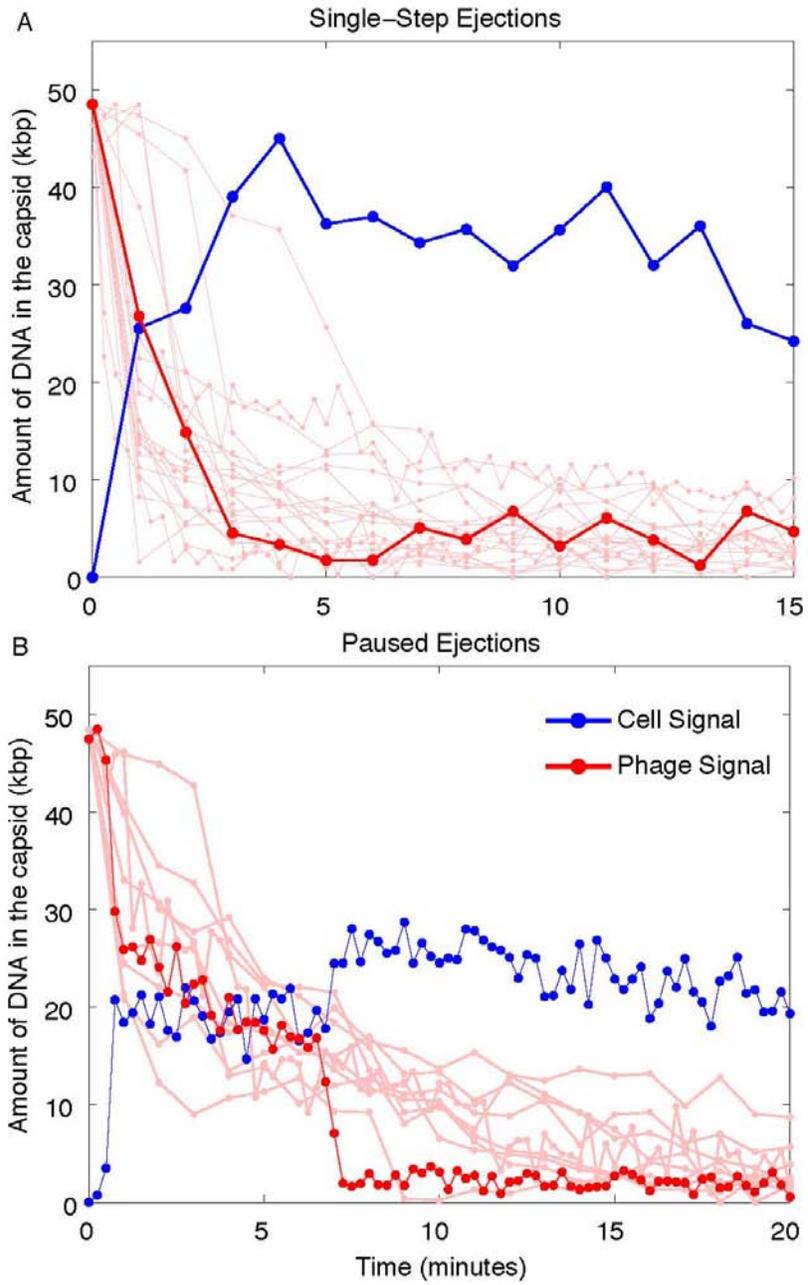

**Figure 3. Ejection trajectories from single-cell infections for λcI60.** The red trajectories show the time history of the DNA intensity within the virus and the blue trajectories show the concomitant increase in the fluorescence in the cellular interior. The solid red color highlights a characteristic ejection and the lighter red color displays other ejection events for reference. The conversion between arbitrary units and kbp was done by first subtracting each trace's minimum observed fluorescence from itself. Each trace was then normalized by the maximum drop in the phage fluorescence and then multiplied by the genome length, which is 48.5 kbp for λcI60. Only two representative traces for the intensity within the cell are shown, with the remaining trajectories available on-line. Multiple lysate preparations from a single stock of CsCl purified phages were used with similar results. A. Trajectories displaying a rapid and continuous ejection. B. Trajectories that exhibit pausing events.

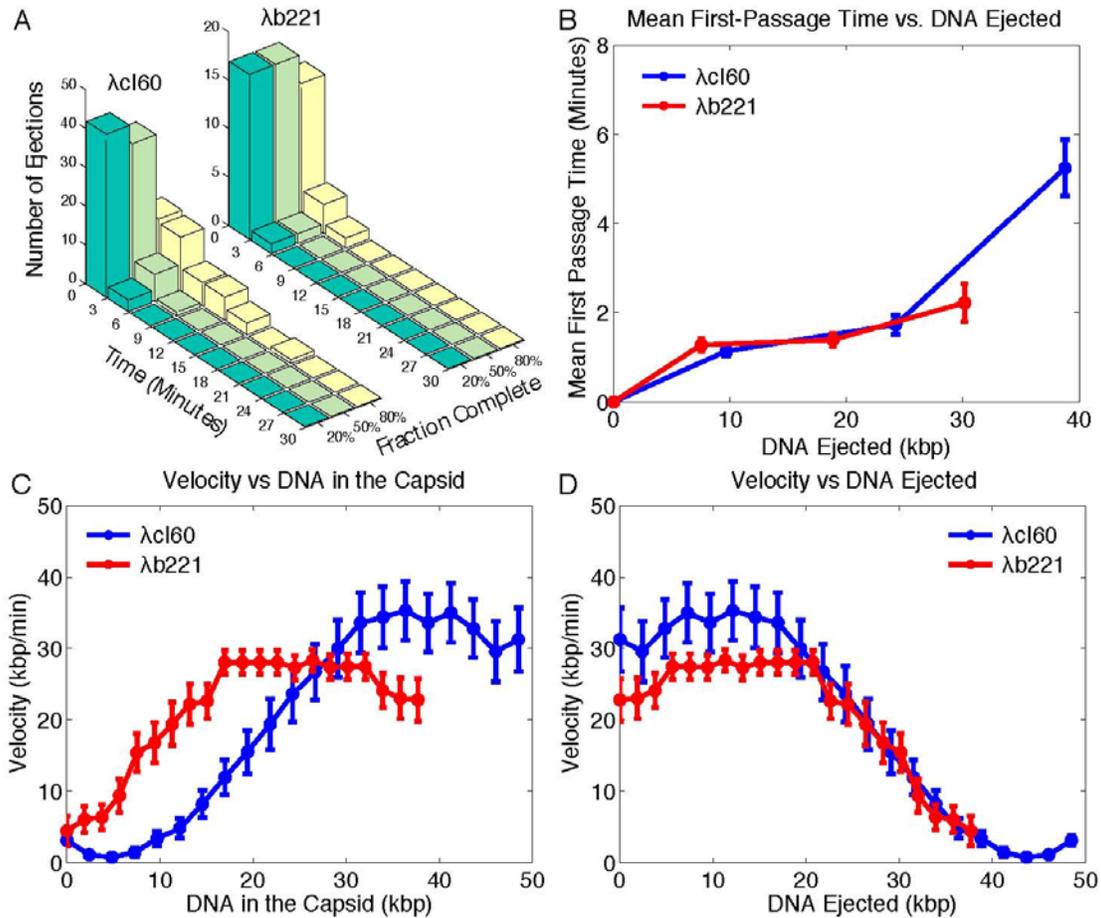

**Figure 4. An ensemble view of ejection times and dynamics for phage λ.** A. Distributions of first-passage times for different fractions of completion of ejection. The histograms are determined by setting ejection thresholds of 20%, 50% and 80% complete as measured using the fluorescence. The distributions for both λcI60 (48.5 kbp) and λb221 (37.7 kbp) are shown. B. Mean first-passage times for λcI60 and λb221. The means of the distributions shown in part A were used to calculate the mean first-passage time. There is little difference in the mean first-passage times of absolute amounts of DNA ejected between λcI60 and λb221. C. Velocity of ejecting DNA plotted as a function of the amount of DNA remaining in the capsid. The initial portion of the ejection process is faster for λcI60 than λb221. There is no significant overlap between the two curves. D. Velocity of ejecting DNA plotted as a function of the amount of DNA ejected. There is a small difference between the two curves for the first 20 kbp, and significant overlap after that. Only trajectories without pauses were used to generate (C) and (D).

Supplemental Information

A Single-Molecule Hershey-Chase Experiment

David Van Valen[*], David Wu[*], Yi-Ju Chen, Hannah Tuson, Paul Wiggins, and Rob Phillips[^]

* Both authors contributed equally to this work.

^ corresponding author  E-mail: phillips@pboc.caltech.edu

Inventory of Supplemental Information

1. Supplemental Figures
    a. Figure S1. Related to Figures 1 and 2. We verified that the SYTOX Orange stain had appropriate specificity and sensitivity to visualize phage lambda.
    b. Figure S2. Related to Figures 1 and 2. We provide evidence that the time scale of the unbinding of SYTOX Orange and concomitant increase in fluorescence in the host cell is consistent with *in vivo* ejections.
    c. Figure S3. Related to Figure 1, 2 and 3. Here we show that SYTOX Orange staining does not affect *in vitro* ejection dynamics. We also demonstrate that SYTOX Orange exchanges quickly with the environment once the phage ejects its DNA, which supports the observation in the paper that the dye redistributes onto the cell genome. We also show that dye leakage occurs on a time scale much longer than ejections. In order to rule out photobleaching artifacts, we measured the rate of photobleaching in our system and demonstrated that the time scales of ejection and photobleaching are well separated.
    d. Figure S4. Related to Figures 2 and 3. We can distinguish ejecting phages from non-ejecting phages that are bound to the same cell.
2. Supplemental Tables. Media and strains used in this work.
3. Supplemental Movies. Examples of phages ejecting into cells.
4. Supplemental Experimental Procedures. Plate lysis. *In vitro* ejection experiments. Data analysis algorithm.
5. Supplemental References

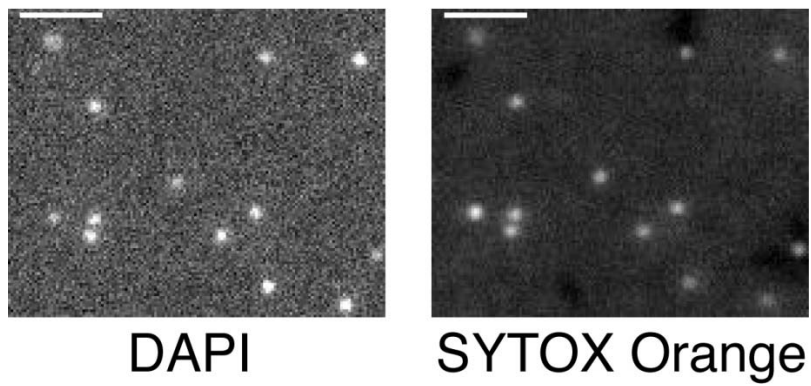

**Figure S1. Simultaneous staining of phage λcI60 with SYTOX Orange and DAPI.** Here, we verify that SYTOX Orange will penetrate the capsid and stain phage DNA; we compared the SYTOX Orange stain with DAPI, which is known to be a quantitative indicator of DNA mass in phages [1]. Phages were stained with a 1.8 mM concentration of DAPI and 500 nM SYTOX Orange, then flowed into an observation chamber. It has been previously established that DAPI will readily penetrate the phage capsid and stain phage DNA [1, 2]. We observe perfect co-localization of the DAPI and SYTOX Orange signals, demonstrating that SYTOX Orange will enter the phage capsid and stain the phage DNA. Staining and co-localization of the viral DNA with both DAPI and STYOX Orange, confirms that SYTOX Orange is both a sensitive and specific indicator for the presence of phage lambda. The presence of fluorescent puncta also provides evidence that SYTOX Orange will not adversely affect phage stability (in contrast to other dyes like SYBR Gold) [3]. The scale bar is 2 microns.

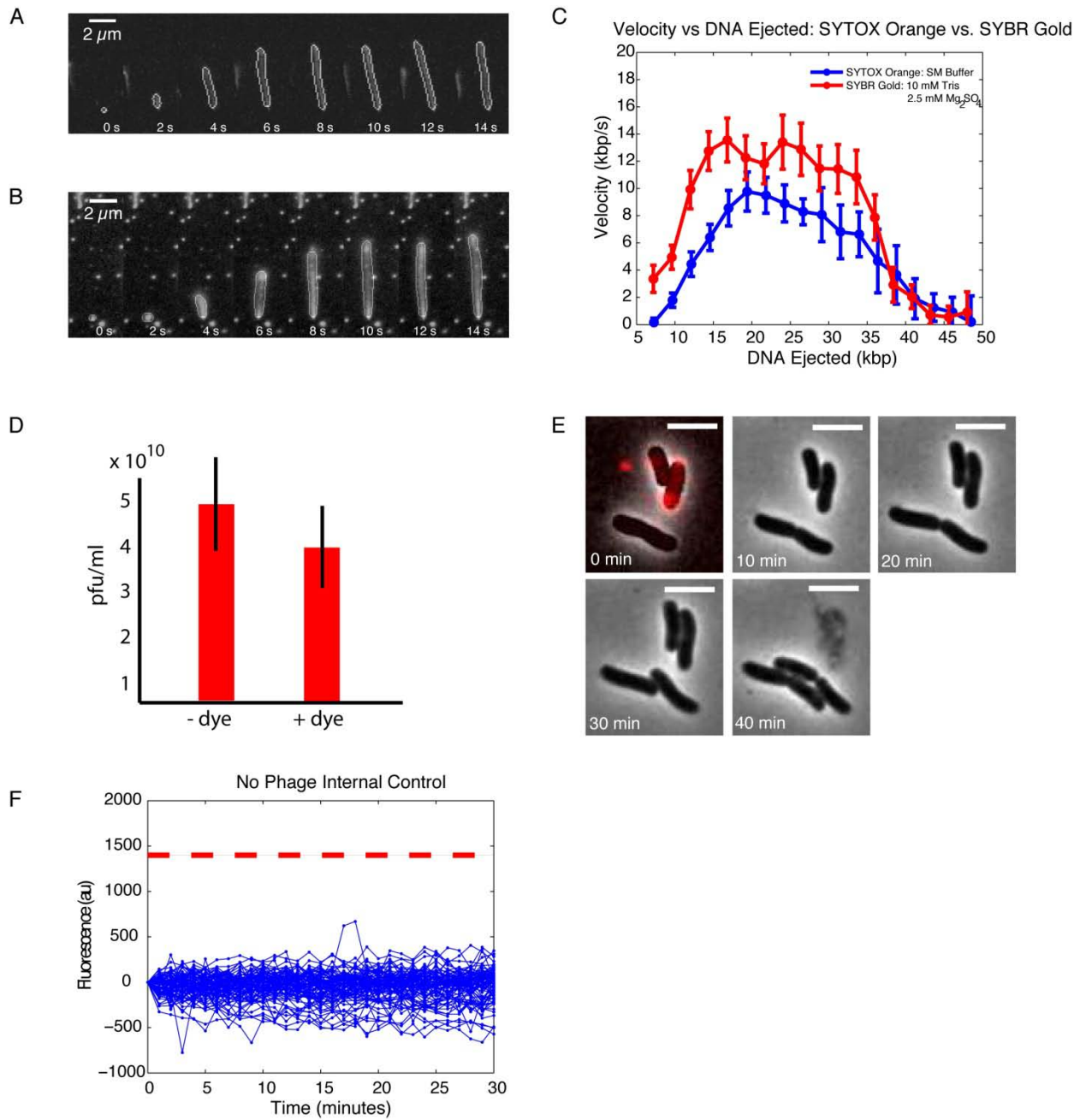

**Figure S2. SYTOX Orange penetrates the phage capsid, maintains phage stability and infectivity, preserves *in vitro* ejection kinetics while stained, and does not cross the membrane of living cells,**
A, B. Data for the single-molecule *in vitro* DNA ejection experiment. An *in vitro* ejection experiment was performed on unstained phages (A) and pre-stained phages (B) to determine whether the presence of dye inside the capsid has any effect on the ejection dynamics. SYBR Gold was used in the experiment shown in (A) and SYTOX Orange was used in the experiment shown in (B). A MATLAB script was used to segment the DNA; the pixels identified as DNA are outlined. C. Velocity vs. the amount of DNA ejected from the capsid for the single-molecule *in vitro* DNA ejection experiment. The presence of dye inside the capsid does not significantly change the underlying dynamics of the ejection process. We hypothesize that difference between the two conditions can be attributed to subtle ionic differences in the buffers for the two experiments. D. We examined the extent to which these dye molecules alter the macroscopic titers of infectious phage. The unstained phage (- dye) had a titer of $5 \times 10^{10}$ while the stained phage (5 µM) (+ dye) had a titer of $3.9 \times 10^{10}$, a drop of 20%. The experiment was performed in triplicate, and the error bars are from counting error statistics, which our experiments obey. The presence of dye inside the capsid does not have a statistically significant effect on the infectivity of lambda phage in bulk. E. A montage demonstrating that cells that have been infected with stained phage will lyse, indicating that presence of the dye neither compromises the infection process nor cellular physiology. SYTOX Orange stained phages are bound to a bacterial cell, placed on an agar pad, and then imaged with fluorescence microscopy over a sufficiently long time to monitor the infection process. Only one fluorescence image was taken to mitigate any possible photo-damage from the excitation of SYTOX Orange, as an oxygen scavenging system was not present. We assume that fluorescence inside a cell signifies that the cell had previously been infected by phage. All of the fluorescent cells that we observed went on to lyse (n=23). Cells with bound phages were also observed; of these, 80% went on to lyse (n=18). This measurement in conjunction with the bulk titering measurement demonstrates that SYTOX Orange does not interfere with the lytic pathway in any substantial way. Scale bar: 4.8 microns. F. No phage control. The fluorescence of 89 cells with no phage attached was monitored over the course of a data set. The red line indicates the average change in fluorescence of phages that ejected within the data set. No increase in fluorescence is seen, demonstrating that SYTOX Orange does not cross the membranes of living cells. Moreover, any fluorescent cells we see we assume have been infected.

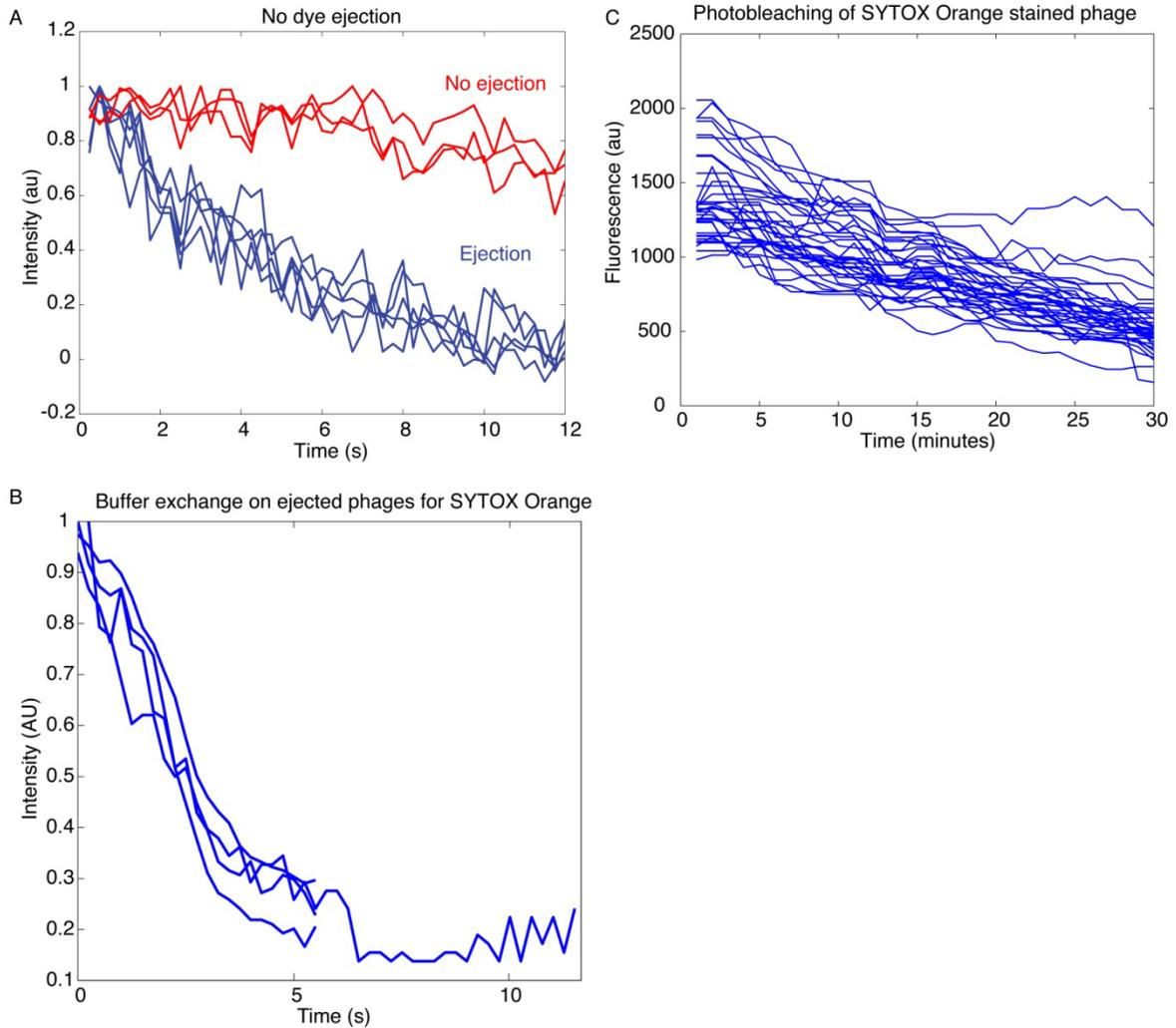

**Figure S3. Single-molecule studies of SYTOX Orange-DNA unbinding kinetics and photobleaching.**
A. *In vitro* DNA ejection kinetics are not affected by SYTOX Orange staining. Dye leakage from the capsid is much slower than ejections. The phage capsids are rapidly destained as DNA is ejected (blue) in the presence of lambda receptor, LamB, and dye falls off. The kinetics are consistent with *in vitro* ejections [4] in which the ejected DNA was stained, instead of the phage capsid DNA, as is shown here. Compare with phage capsids which did not eject (red) which demonstrates a drop in fluorescence intensity presumably due to photobleaching. The kinetics are markedly different. These ejection curves demonstrate that the *in vitro* kinetics of ejection are unaffected by staining the phage capsid DNA with SYTOX Orange. B. Single-molecule *in vitro* DNA ejection in the presence of SYTOX Orange. We performed an *in vitro* ejection experiment with SYTOX Orange instead of SYBR gold following the method in [3] with buffer exchange. Phages affixed to a coverslip are made to eject into a buffered solution containing SYTOX Orange and we wait for the ejection to complete. Then at t = 0 in the figure, the buffer is exchanged with a solution without any dye; as is shown above, the fluorescence signal drops by 70% within 7 seconds. From [5], the $k_{off}$ for SYTOX Orange is 0.58 s$^{-1}$, and they achieve a 38% reduction in staining with a 100-fold dilution of 500 nM SYTOX Orange within a few seconds, which is consistent with our observations. In our view, these experiments support the interpretation of the *in vivo* ejection assay that when the viral genome enters the cell, the bound dyes fall off and they can then bind onto the much larger host cell genome, allowing us to visualize intracellular fluorescence. C. Dye leak and photobleaching of SYTOX Orange stained bacteriophage. In order to properly characterize the ejection events, the combined effect of passive loss of dye from phage and photobleaching must be distinguishable from viral DNA ejection. Phages were stained with SYTOX Orange (as described in the methods section), flown into a KOH cleaned flow chamber, and then allowed to non-specifically adhere to the glass surface. The phages were then imaged with a frame rate of 1 min$^{-1}$, an exposure time of 500 ms, and with a TRITC filter. These imaging conditions mimic the conditions used to collect most of our data. The trajectories of the total fluorescence above background for 36 phages are shown. The time scale for dye loss and photobleaching is 30 minutes, and all the trajectories are monotonically decreasing. While significant loss of signal does occur over the course of 30 minutes, ejections, as shown in Figures 3D and 4 in the main text, range from 1 minute to 20 minutes. Thus for the typical ejection time scale of 15 minutes, dye loss and photobleaching can account for at most a 30% loss of signal. On the other hand, the fluorescence losses seen for the putative ejection events were much more stereotyped and typically faster than the monotonic decreases seen to result from explicit photobleaching. This lends credence to our use of the rapid decrease in phage signal as a marker for putative *in vivo* ejection events.

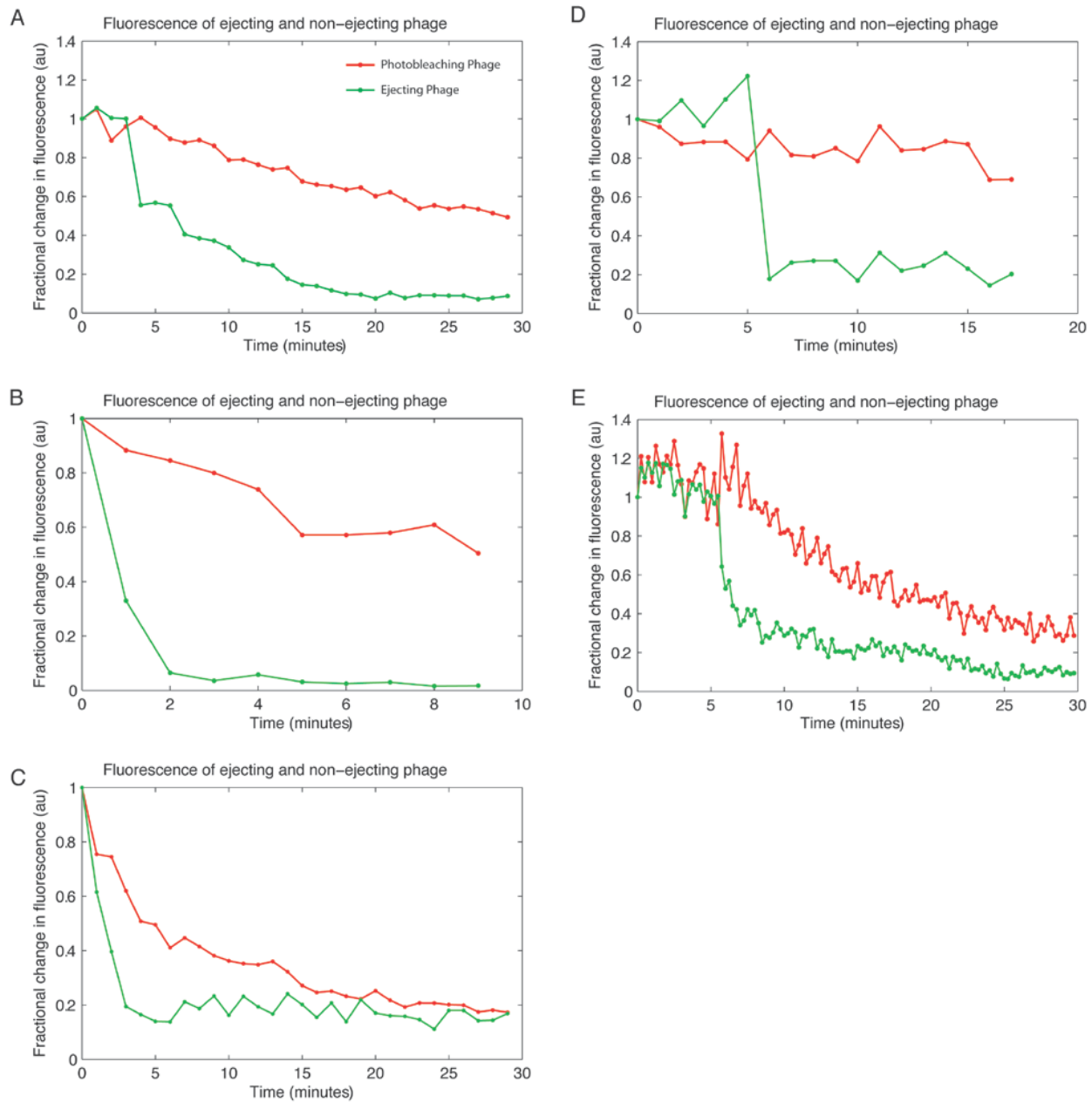

**Figure S4. Photobleaching in SYTOX Orange stained bacteriophage bound to cells.** For those cases in which only one of the bound phages underwent a putative ejection event, by simultaneously monitoring the level of fluorescence in the other, non-ejecting phage, we could directly compare an ejecting and non-ejecting phage, both subject to photobleaching. Some of these phages did not display a significant decrease in fluorescence, indicating that they did not infect the cell. Here we plot the decrease in fluorescence in an infecting phage and a non-infecting phage in five different cells (A--E). The green lines were determined to be ejecting phages, because their fluorescence decrease mirrors the increase in fluorescence inside the cell, and is much faster than photobleaching. The red lines, therefore, represent the fluorescence of non-ejecting phages. The total fluorescence above background for each phage was determined using segmentation masks as demonstrated in Figure 2 in the main text. These values are normalized by the maximum observed fluorescence in each phage to allow for side-by-side comparison of the drop in fluorescence levels. The data demonstrates that the loss in fluorescence in ejecting phages occurs on a faster time scale than photobleaching and dye loss, which allows us to distinguish ejection from photobleaching.

## Strains and media

Here, we detail the stains that were used during this work (Table S1) and recipes for media and buffers (Table S2).

| Strain, Phages, or Plasmid | Description |
|---|---|
| LE392 (Promega, K9981) | "Wild-type" *E. coli* that carries the supF amber suppression phenotype, allowing Sam7 phages to undergo lysis |
| λcI60 (Gift from Michael Feiss, University of Iowa) | A lambda phage with a clear plaque morphology mutation in the cI gene (48.5 kbp genome) |
| λb221 (Gift from Michael Feiss, University of Iowa) | A lambda phage with a deletion in the b region (37.7 kbp genome) |

**Table S1.** Strain and plasmids that were used or constructed during the course of our study.

| Media | Description |
|---|---|
| NZYM (Teknova, N0170) | 10g NZ amine, 5g NaCl, 5g Bacto-yeast extract, 2g $MgSO_4$ -$7H_2O$, 1 L milliQ water, autoclave |
| LB (EMD Millipore, 71753-5) | 10g Bacto-tryptone, 5g yeast extract, 10g NaCl, 1 L milliQ water, autoclave |
| LBM | LB + 10 mM $MgSO_4$-$7H_2O$ |
| M9sup | 1X M9 salts, 1 mM thiamine hydrochloride, 0.4% glycerol, 0.2% casamino acids, 2 mM $MgSO_4$, 0.1 mM $CaCl_2$ |
| M9maltose-sup | 1X M9 salts, 1 mM thiamine hydrochloride, 0.4% maltose, 0.2% casamino acids, 2 mM $MgSO_4$, 0.1 mM $CaCl_2$ |
| SM | 5.8 g NaCl, 2 g $MgSO_4$-$7H_2O$, 50 mL 1 M Tris-Cl, pH 7.5, 2 % w/v gelatin (BD, 214340), milliQ water to 1 L, autoclave |
| TM | 50 mM Tris-HCl (pH 7.4), 10 mM $MgSO_4$ |
| GODCAT | 100 nM glucose oxidase (from *Aspergillus niger*, Sigma, G2133), 1.5 µM catalase (from bovine liver, Sigma, C1345), 500 µL TM |

**Table S2.** Recipes for various media and buffers used in our study.

## Movies

Characteristic videos of *in vivo* DNA ejection (Movies S1, S2, and S3) are available online. These movies show the ejection highlighted in Figure 3 (Movie S1), a single-step ejection (Movie S2), and a paused ejection (Movie S3).

## Experimental Procedures

**Plate lysis**. For a detailed description of components and strains, please see Tables S1 and S2 in the SI. NZYM top agarose (NZYM + 0.7% agarose) and NZYM plates (NZYM + 1.5 % agar (BD, 214010)) were prepared prior to plate lysis. NZYM top agarose was melted on a hot plate and then stored in a 45 °C water bath until needed. The host cell strain, LE392, was grown up overnight in 5 mL of LB. The saturated cell culture was then centrifuged for 5 minutes at 5,000 g and the pellet was resuspended in 5 mL of SM buffer. A 100 µL aliquot of cells was then mixed with 1 µL of phage stock in a 14 mL culture tube and incubated at 37 $^{o}$C for 20 minutes. Next, 3 mL of NZYM top agarose was added to the culture tube, gently mixed, and poured onto a NZYM plate. The plates were incubated for 12--16 hours at 37 $^{o}$C or until lysis was visually apparent. After incubation, phages were recovered by pouring 5 mL of SM buffer onto the plate and placing the plate on a rocking station at 4 $^{o}$C. After 5 hours, the SM buffer was recovered. The lysate was sterilized by adding chloroform to a concentration of 1% and gently vortexing. Bacterial debris and chloroform were then removed by centrifuging for 10 minutes at 5,000 g; the supernatant was recovered. Plate lysis typically yielded titers of ~ $10^{10}$ pfu/ml. CsCl purification of phages was performed as in [].

**Real-time imaging of DNA ejection** *in vitro*. We follow a protocol that was first developed by Mangenot *et al.* and later adapted to use with phage lambda [3, 4, 6]. Microscope coverslips (18x18 mm, #1.5, VWR, 48367-092) were cleaned by sonication in 1M KOH for 10 minutes followed by sonication in water for 10 minutes and dried on a hot plate. Glass slides (75 x 25 mm, VWR, 48300-263) were drilled using a diamond covered drill bit and 5 inches of tubing was attached to the glass slide using epoxy. The flow chamber was assembled using laser cut double-sided adhesive tape (Grace Biolabs, SA-S-1L). A solution of $10^{8}$-$10^{11}$ pfu/ml lambda phage was incubated in the assembled flow chambers for 10 minutes. Once focused, the chamber was washed with 200 µL of buffer + 1% oPOE (Alexis Bio-chemicals, 500-002-L005). Buffers were either SM buffer for the SYTOX Orange (Invitrogen, S11368) measurement or 10 mM Tris, pH7.5, 2.5 mM $MgSO_4$. The solution to induce ejection consisted of buffer, 1% oPOE, 1% glucose oxidase/catalase, 1% LamB, 0.5% glucose, 1% beta-mercaptoethanol, and either $10^{-6}$ diluted SYBR Gold (Invitrogen, S11494) or 500 nM SYTOX Orange. Calibration of lengths and data analysis was performed as in [3]. LamB was extracted from the membranes of *E. coli* pop154 cells [3]; these cells express a lamB gene from *S. sonnei*.

**Image analysis.** Cells of interest were manually identified in each movie and cropped from the field of view using ImageJ. The phase images for each cell were segmented using custom image analysis software created with MATLAB. Briefly, the phase images from each time point in the movie were registered with the first frame by cross-correlation; this registration was then applied to all fluorescence channels, removing spatial drift from the data set. Next, the registered phase images were all added together; this step greatly reduces difficulty of segmentation. Lastly, thresholding and minor morphological operations were used to create a mask of the cell from the combined image. Cells were occasionally segmented manually when quality phase images were not available. Phage masks were created from the fluorescence channel by a similar process. Quantities of interest, including background levels, and fluorescence intensities inside the cell and phage in each frame were extracted for further analysis.